\documentclass[10pt,conference]{IEEEtran}
\usepackage{graphicx}
\usepackage{xcolor}
\usepackage{comment}

\ifCLASSINFOpdf
\else
\fi
\usepackage{listings}

\usepackage{graphicx}

\definecolor{codegreen}{rgb}{0,0.6,0}
\definecolor{codegray}{rgb}{0.5,0.5,0.5}
\definecolor{codepurple}{rgb}{0.58,0,0.82}
\definecolor{backcolour}{rgb}{0.95,0.95,0.92}

\lstdefinelanguage{openlinealgorithm}{
morekeywords={begin,end,procedure,if,then,else,for,in,repeat,until,break,when,or,mandatory,or_inclusive,or_exclusive,optional},
morecomment=[l]{--}
}

\lstdefinestyle{openlinestyle}{
  backgroundcolor=\color{backcolour},   
  commentstyle=\color{codegreen},
  keywordstyle=\color{blue}\textbf,
  numberstyle=\tiny\color{codegray},
  stringstyle=\color{codepurple},
  basicstyle=\ttfamily\footnotesize,
  breakatwhitespace=false,         
  breaklines=true,                 
  captionpos=b,                    
  keepspaces=true,
  numbers=left,                    
  numbersep=5pt,                  
  showspaces=false,                
  showstringspaces=false,
  showtabs=false,                  
  tabsize=2
}

\usepackage{url}
\usepackage{flushend}

\begin{document}
\title{Onboarding in Software Product Lines: Concept Maps as Welcome Guides}

\author{\IEEEauthorblockN{Maider Azanza}
\IEEEauthorblockA{University of the Basque\\Country (UPV/EHU)\\
San Sebasti{\'a}n\\
Spain\\
maider.azanza@ehu.eus}
\and
\IEEEauthorblockN{Arantza Irastorza}
\IEEEauthorblockA{University of the Basque\\Country (UPV/EHU)\\
San Sebasti{\'a}n\\
Spain\\
arantza.irastorza@ehu.eus}
\and
\IEEEauthorblockN{Raul Medeiros}
\IEEEauthorblockA{University of the Basque\\Country (UPV/EHU)\\
San Sebasti{\'a}n\\
Spain\\
raul.medeiros@ehu.eus}
\and
\IEEEauthorblockN{Oscar D{\'i}az}
\IEEEauthorblockA{University of the Basque\\Country (UPV/EHU)\\
San Sebasti{\'a}n\\
Spain\\
oscar.diaz@ehu.eus}
}


%Telephone: (800) 555--1212, Fax: (888) 555--1212}


\maketitle

\begin{abstract}

With a volatile labour and technological market, onboarding is becoming increasingly important. The process of incorporating a new developer, a.k.a. the newcomer, into a software development team is reckoned to be lengthy, frustrating and expensive. Newcomers face personal, interpersonal, process and technical barriers during their incorporation, which, in turn, affects the overall productivity of the whole team. This problem exacerbates for Software Product Lines (SPLs), where their size and variability combine to make onboarding even more challenging, even more so for developers that are transferred from the Application Engineering team into the Domain Engineering team, who will be our target newcomers. This work presents concept maps on the role of sensemaking scaffolds to help to introduce these newcomers into the SPL domain. Concept maps, used as knowledge visualisation tools, have been proven to be helpful for meaningful learning. Our main insight is to capture concepts of the SPL domain and their interrelationships in a concept map, and then, present them incrementally, helping newcomers grasp the SPL and aiding them in exploring it in a guided manner while avoiding information overload. This work's contributions are four-fold. First, concept maps are proposed as a representation to introduce newcomers into the SPL domain. Second, concept maps are presented as the means for a guided exploration of the SPL core assets. Third, a feature-driven concept map construction process is introduced.  Last, the usefulness of concept maps as guides for SPL onboarding is tested through a formative evaluation.

Link to the online demo: \url{https://rebrand.ly/wacline-cmap}
%State-of-the-art SPL models document variability and the domain in an extensive fashion, but are too complicated. Newcomers usually experience information overload using these models. 
\end{abstract}


% \fi
\IEEEpeerreviewmaketitle

\maketitle

\section{Introduction}
% \cite{Balali18, Begel08, Heimburger2020, Pham17, Pradel16, Rastogi15, Rastogi17, Steinmacher14, Viviani19}
Onboarding refers to the process that arises when incorporating new developers into a software development team \cite{Pham17, Rastogi17, Steinmacher14, Viviani19}. This process is often frustrating for all parties. Project managers must endure reduced productivity and sacrifice senior developers' time to mentor the newcomer. Senior project members are required to pass on a large amount of information in a short period. Finally, newcomers are faced with the challenge of familiarising themselves with an entirely new work environment, with the awareness that they are occasionally keeping their senior colleagues from their work by asking for guidance \cite{Pradel16}. These problems exacerbate for \textit{Software Product Lines (SPLs)}.

%MA: Add a brief intro to SPLs including their advantages, as reviewers might not be familiarized with it.
An SPL is a set of software-intensive systems sharing a common, managed set of features that satisfy the specific needs of a particular market segment or mission and that are developed from a common set of core assets in a prescribed way \cite{Clements01}. Two issues play a crucial role in SPLs. The first issue is adequately handling variability, which is the ability to derive different products from a common set of core assets. The second issue is how to systematically build such core assets so that they will later be reused to yield different products \cite{Apel13}. SPLs are typically developed using two processes, \textit{Domain Engineering} and \textit{Application Engineering}. The former analyses the domain of the SPL and develops the aforementioned core assets, that is, software artefacts such as code, architecture, requirements, and so on, that will serve as building blocks for the SPL products. In this sense, Domain Engineering targets \textit{development for reuse}. On the other hand, Application Engineering has the goal of developing a specific product or application for the requirements of a particular customer by reusing core assets. That is, it targets \textit{development with reuse}. The notion of feature plays a central role in both processes, as features encapsulate requirements' variability in terms of elements the customer can choose from to yield a product, and they are also used to specify how such requirements are realised in the core assets. SPLs have shown measurable benefits such as reduced time-to-market, reduced costs, or increased quality\footnote{Available at: \url{https://splc.net/fame.html}} \cite{Apel13}.

% Issues with SPL onboarding
While the onboarding process into application engineering teams bears a resemblance to onboarding into single system development teams, SPLs pose a unique set of challenges to newcomers that are incorporated into the domain engineering team. Even if we set their sheer volume aside, where the same SPL can yield thousands of products, both handling variability and developing for reuse complicate onboarding. First, feature models, the hallmark artefact to specify variability, can gather together multiple concerns, leading to an increase in complexity and making them hard to comprehend and maintain \cite{Capilla13}. On top of this, if we analyse how this variability is carried out in the core assets, we encounter that preprocessor directives (\#ifdef's), which are often used to implement compile-time variability, have been reported to impact code comprehensibility negatively \cite{Ernst02, Melo17}, as the functionality for each feature is distributed in different core assets and intertwined with other functionalities. As a consequence, it is difficult to grasp the concept of large-scale reuse \cite{Acher17}. The results of these challenges also exceed those that arise when onboarding occurs in single system development teams. In the latter case, an error introduced by a newcomer impacts only one system, in domain engineering it has the potential to affect all products that are built using that particular core asset.

% Relevance of SPL onboarding
Yet, onboarding is particularly crucial for the future of an SPL. Not only do SPLs face the need to incorporate people from outside, but in-house migration might also occur. On the way to become fully configurable product families, where every product is automatically created from core assets and application engineering is no longer necessary, SPLs experiment a movement of human resources from the application engineering team to the domain engineering team \cite{Deelstra04}. That is the reason why our work focuses in this scenario, where the newcomer is familiar with one or more products of the SPL but needs to understand the variability and the development \textit{for reuse} required to work in the core assets.

To the best of our knowledge, there is no previous work that addresses onboarding in SPL domain engineering teams. For single-system development, typical strategies to tackle onboarding include courses \cite{Pradel16, Sim98}, bootcamps \cite{Pham17} or mentors \cite{Balali18, Pham17}, where senior developers are appointed to guide the newcomer in this process. However, these strategies are costly in terms of both time and money, and appointing senior developers as mentors can impact productivity, which can especially hurt small teams \cite{Pham17}. As a result, most development teams expect newcomers to explore and understand the source code by themselves \cite{Viviani19}. 

% Our proposal
Nevertheless, directly exploring the SPL is not feasible for newcomers. Main stumbling blocks include the above-mentioned size, understanding variability and internalising development for reuse \cite{Acher17}. Needed is a way to provide a view at a higher abstraction level so that the newcomer can acquire a perspective of the SPL as a whole. To this end, we advocate for building \textit{sensemaking scaffolds} \cite{Quintana04} on top of the SPL so that the newcomer can explore the SPL independently, but in a guided manner.

Specifically, we propose \textit{concept maps} \cite{Novak08} as the means to create such scaffolds. Concept mapping is reckoned to be a means for meaningful learning insofar as it serves as a kind of template or scaffold to help to organise knowledge \cite{Nesbit2006}. On these premises, we explore the following research questions:

\begin{itemize}
\item RQ1: Are concept maps adequate sensemaking scaffolds to ease onboarding in SPL domain engineering teams? (Section \ref{sec:SensemakingConceptMaps})
\item RQ2: How does an SPL concept map look like? (Section \ref{Sec_SPL_CMaps})
\item RQ3: How is an SPL concept map constructed? (Section \ref{Sec_construction})
\end{itemize}

We elaborate on these questions and present a pilot study for the \textit{WACline SPL} (Section \ref{Sect_Evaluation}). We start by introducing the onboarding problem.

\section{The Onboarding Problem} \label{Sec_TheOnboardingProblem}
Newcomers face different types of barriers when they are incorporated into a new software development project \cite{Balali18}, namely:
\begin{itemize}
\item \emph{personal barriers}, which include newcomers' reluctance to ask for help from their colleagues early in their problem-solving processes for fear of wasting their time \cite{Balali18, Begel08, Pradel16}, 
\item \emph{interpersonal barriers}, which, as a case in point, refer to communication issues that arise when newcomers are incorporated into a diverse team, where different people with different goals, different cultures and different interpersonal skills gather together \cite{Balali18},
\item \emph{process barriers}, where newcomers encounter difficulties in having a holistic perspective of the software they are there to contribute to, and in finding where to start working \cite{Steinmacher14, Viviani19},
\item \emph{technical barriers}, newcomers often encounter problems due to the high complexity of the systems being developed \cite{Balali18, Steinmacher14}. This problem is accompanied by the lack of prior knowledge of the domain where development occurs \cite{Steinmacher14}.
\end{itemize}

These barriers impact not only newcomers, but also the other stakeholders involved in the process. Newcomers experience frustration \cite{Pradel16, Sim98}. Senior developers struggle with their new mentoring role \cite{Balali18}. Finally, project managers must endure reduced productivity and sacrifice senior developers' time to mentor the newcomer \cite{Pradel16}. As described in the introduction, the onboarding process is costly both in time and money for an organisation. This is the reason why adding personnel to a project actually decreases productivity in the short term \cite{Sim98}.

In most cases, newcomers explore the source code by themselves \cite{Viviani19}. This might or might not be accompanied by a rich documentation. However, just providing a bunch of documentation leads to information overload \cite{Steinmacher14} and the same can be said for specific onboarding sessions, where, while newcomers find the information useful for context, they are often overwhelmed by the amount of information they receive and sometimes struggle to understand the pertinence of the information to their job \cite{Viviani19}.

While the barriers above refer to software development teams in general, these issues are also applicable or even aggravated in the case of the onboarding process in the domain engineering team of SPLs. Note that domain engineering involves more complex artefacts and dealing with both variability and development for reuse. These reasons lead us to focus on the technical barriers. Furthermore, new members need a good command of the problem space (i.e. the domain), as the foundation on which to build knowledge of the solution space (including architecture, code, and so on). At this point, it is important to clarify what \textit{domain}, as in domain engineering team, refers to. 

According to Apel et al. \cite{Apel13} in the SPL realm, a \textit{domain} is an area of knowledge that:  
\begin{itemize}
    \item is scoped to maximise the satisfaction of the requirements of its stakeholders,
    \item includes a set of concepts and terminology understood by practitioners in that area,
    \item and includes the knowledge of how to build software systems (or parts of software systems) in that area, that is, the products that comprise the SPL.
\end{itemize}

If the SPL is in place, the domain scope has already been worked out. Hence, we focus on the last two issues. That is, first, we focus on helping newcomers understand the set of concepts and terminology of the area and, second, on helping them acquire the knowledge of how to build the SPL products, that is, understand variability and how it is realised in the core assets that are developed \textit{for reuse}. 

The following section frames our work with respect to previous work in onboarding for single-system development. 

\section{Related Work}

Shortening and facilitating onboarding has raised considerable interest in software engineering \cite{CherryABR04,Heimburger2020,Valerio06} in general, and in the area of \textit{Open Source Software (OSS)} in particular \cite{CanforaPOP12,ParkJ09,SteinmacherCTG16,SteinmacherGCR19,WangS11,CubranicMSB05}. Yet, as far as we know, no previous work tackles the impact of SPL specifics in onboarding. Thus, this section concentrates on works about onboarding in software development teams in general. We study them along the following comparison framework: (1) \textit{context}, i.e. the environment in which the onboarding takes place; (2) \textit{population}, i.e. newcomers' profile; (3) \textit{intervention}, i.e.  type of action to act upon onboarding; and (4), \textit{outcome}, i.e. reported impact. 

\textbf{Context}. So far, most studies are balanced between onboarding in industrial contexts \cite{Heimburger2020, Sim98, Pham17, Viviani19} and in open source projects \cite{ParkJ09, SteinmacherCTG16,Panichella15, Balali18}. Our work is targeted to onboarding in domain engineering teams of SPLs, a context that has not yet been studied.

\textbf{Population}. Newcomers might come from different backgrounds: freshly graduated newcomers \cite{Pham17}; self-paced learners  \cite{Hibschman19}, generations Y or Z \cite{Heimburger2020} or practitioners \cite{Sim98,Viviani19,Yates19}. Other works do not specifically profile their target audience, but just talk about newcomers  \cite{CubranicM03, Dagenais10, ParkJ09, SteinmacherCTG16, Panichella15, WangS11, MalheirosMTM12}. Our work focuses on newcomers that are incorporated into the domain engineering team from the application engineering team. 

\textbf{Intervention}. Easing and speeding up onboarding has been the subject of distinct interventions: organising specific courses or bootcamps for newcomers \cite{Balali18, Pham17, Sim98,Viviani19, Yates19}; establishing a buddy system \cite{Viviani19}; setting a gamification system \cite{Heimburger2020}; tools for assisted coding  \cite{Balali18, Hibschman19, Pham17, Sim98, Viviani19}; visualisation of data and conceptual representations \cite{Hibschman19, ParkJ09}; recommendation assistants \cite{CubranicM03, Panichella15, MalheirosMTM12}. Specifically, the study of Park and Jensen \cite{ParkJ09} shows how code visualisation tools can speed up the learning curve and help newcomers find information faster and more effectively. Malheiros et al. \cite{MalheirosMTM12} and Cubranic et al. \cite{CubranicM03,CubranicMSB05} present two tools that support newcomers by selecting the most appropriate source code files for their development tasks. \textit{Mentor} \cite{MalheirosMTM12} makes recommendations based on files changed in similar tasks, while \textit{Hipikat} \cite{CubranicM03,CubranicMSB05} builds a project memory from where newcomers can make queries that fit their tasks. Our work also proposes a visualisation mechanism, tailored to domain in SPLs. Specifically, it should help the newcomer learn the set of concepts and terminology understood by practitioners and should incorporate the knowledge about the products of the SPL.

\textbf{Outcome}. Regarding the results of the previous interventions, some works mention the improved subjective experience of newcomers \cite{Heimburger2020, Yates19, Viviani19}. However, these works analyse interventions such as establishing a buddy system, specific courses or a gamification system. If we focus on proposals of visualisation mechanisms such as ours, Park et al. \cite{ParkJ09} underline the improved experience, and also the efficiency in learning tasks:  \textit{"Even novices with little training were able to benefit from these tools. Providing more efficient ways to handle large amounts of information can lower the learning curve and information overload newcomers experience when joining an OSS project"}. The \textit{Isopleth} tool \cite{Hibschman19} also helps junior developers improve their understanding of individual code components, relationships among components, and the dataflow. It provides support for newcomers to use that built understanding to further their exploration. In the same sense of being entry gates for further tasks, Cubranic and Murphy \cite{CubranicM03} agree: \textit{"the recommendations ... are still just entry points. The developer must still evaluate the code and understand how it works"}. Last, Hibschman et al. introduce another interesting observation: the advice of letting newcomers progress at their own pace \cite{Hibschman19}. 

We can now better profile our aim: investigating the role of concept maps (Intervention) in SPL domain engineering onboarding (Context) for application engineers (Population) to lower the learning curve and to allow them learn at their own pace (Outcome).  Main requirements include avoiding information overload (preventing frustration) and self-paced learning (reducing mentors' intervention). 

% Lista de referencias utilizadas en el related work, en el orden en que están en la tabla de Framendley
% \cite{Heimburger2020, Hibschman19, CubranicM03, Dagenais10, ParkJ09, SteinmacherCTG16, Panichella15, WangS11, MalheirosMTM12, Pirolli05, Sim98, Pham17, Balali18, Yates19, Viviani19} 

\section{Concept Maps as Sensemaking Scaffolds for SPLs}
\label{sec:SensemakingConceptMaps}

Sensemaking refers to the process of building understanding by generating representations that explain what is known or understood \cite{Weick95}. Sensemaking tackles the challenges learners may face in making sense of examples and artefacts, not only for seeking information but also for building conceptual knowledge about a domain. As a case in point, in software engineering, this theory has been used to help inexperienced developers make sense of professional Web applications \cite{Hibschman19}. Thus, scaffold structures (through software or by other means) are offered to the learners so that they can lean on them in their own sensemaking process.

This aligns with our goal: facilitating newcomers' sensemaking of the SPLs. Specifically, as newcomers are being incorporated into the domain engineering team, domain understanding becomes a cornerstone. Implications are twofold: understanding the concepts and terminology used by practitioners in the area, and understanding how to build the products of the SPL \cite{Apel13}.

Learning sciences' literature provides guidelines to build sensemaking scaffolds \cite{Hibschman19, Quintana04}. Succinctly:
\begin{itemize}
    \item organise tools and artefacts around the semantics of the discipline,
    \item use representations and language that bridge learners' understanding, 
    \item use representations that learners can inspect in different ways to reveal important properties of the underlying data.
\end{itemize}

Along the first guideline, we propose \textit{concept maps} as the central artefact around which the sensemaking scaffold is built (see Fig. \ref{fig:conceptmap}\footnote{As an example of how our proposal can be realised, the concept map is created using \textit{CMapTools}, available at \url{https://cmap.ihmc.us/}}). Concept maps are graphical tools for organising and representing knowledge \cite{Novak08}. They include concepts, usually supported as nodes, and relationships between concepts, indicated by a connecting line linking two nodes. Concept mapping is reckoned to be a means for meaningful learning as it serves as a scaffold to help organise knowledge \cite{Nesbit2006}. In our realm, concept maps can help newcomers understand the main concepts and terminology used by practitioners in the field, where the map gathers the principles that describe the domain for which the SPL has been built. Moreover, concept maps can also be reviewed more quickly than other forms of text-based documentation. 

Along the second guideline, the representation should bridge learners' understanding. Ausubel’s cognitive psychology conjectures that learning takes place by the assimilation of the new cognitive model into the existing cognitive model of the recipient \cite{ausubel1963psychology}. In this sense, "the existing cognitive model" refer to newcomers' existing knowledge obtained through their previous involvement in one or more products of the SPL. Concept maps can be created for both products and the SPL itself. By contrasting newcomer's  product concept map vs. the SPL concept map,  newcomers can more effectively identify the concepts (i.e. nodes) they are not familiar with and, what is even more important, identify how the known and new concepts relate to each other. The concept map provides transitioning paths during the onboarding journey.

Finally, the third guideline refers to using representations that learners can inspect in different ways. This is where concept maps by themselves fall short. In SPL domain engineering onboarding, these ways might refer to distinct abstraction levels at which the core assets reside (e.g. feature diagrams or requirements reside at the problem realm, while the architecture or code reside at the solution realm). Hence, we propose that concept maps act as hubs that help newcomers transit between them. In this sense, concepts will trace together the different core assets, so that each concept will exhibit the SPL features for which it is relevant, and different links will allow the newcomer to explore the assets related to that feature, such as the feature diagram documentation, an architecture component, or the codebase.

\section{Running Example: WACline}\label{Sec_wacline}

Throughout the rest of the paper, we will use the WACline SPL as the running example to illustrate our approach. WACline (\textit{Web Annotation Clients software product line})\footnote{WACline is available at: 
\url{https://onekin.github.io/WacLine/}} is an academic SPL that handles variability in creating different Web annotation browser clients. Currently, WACline holds over 16000 LOC. from which 6585 are variable, it is implemented in JavaScript, and it uses \textit{pure::variants} as the variability management tool \cite{Beuche2019} and \textit{Git} as the version control system.

Web Annotations recreate and extend traditional annotations (i.e. marginalia, errata, and highlights) as a new layer of interactivity and linking on top of the Web \cite{w3crecommendation}. In 2017, the World Wide Web Consortium (W3C) published its recommendations and working group notes for Web Annotation technologies \cite{w3crecommendation}. One of its specifications is the Web Annotation Architecture, which is constituted by two constructs: (1) Annotation Server, that makes annotations available and allows their management, and (2) the Annotation Client, that establishes communication channels with the Annotation Server. Within this context, the aim of WACline is to yield Web Annotation Clients adapted for specific goals in different domains (e.g. exam marking in education, peer review in research, etc.). 

\begin{figure*}
  \centering
      \includegraphics[width=\textwidth]{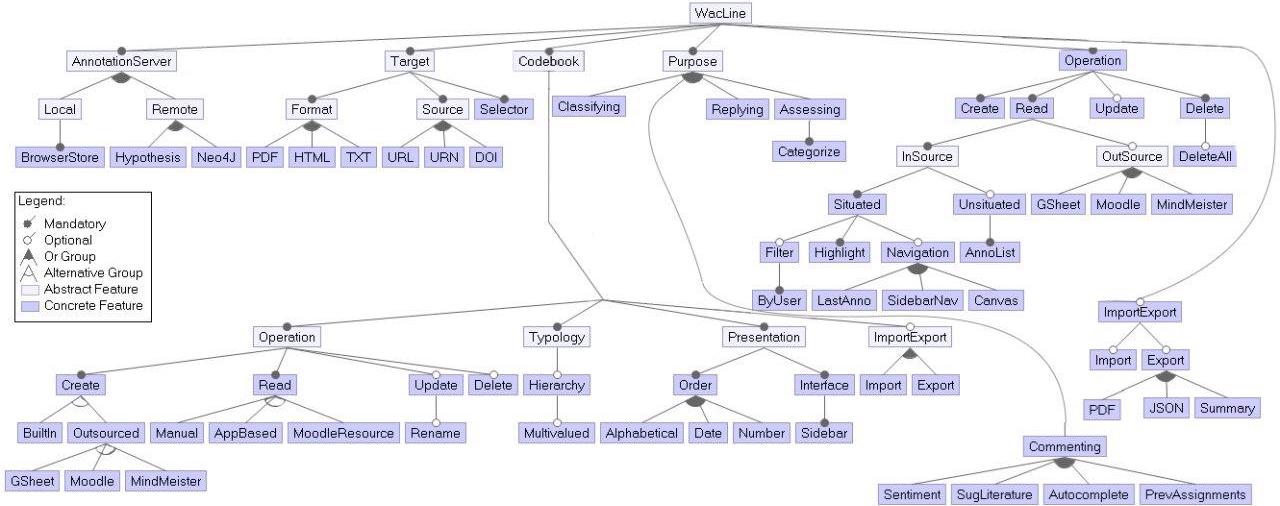}
  \caption{Feature Diagram of WACline.}
  \label{fig:featurediagram}
\end{figure*}

WACline annotations comprise a \textit{target} (i.e. where the annotations have been defined), a \textit{purpose}  (i.e. the rationale for the annotation), etc, and can be manipulated through \textit{operations}. All these aspects exhibit variability, e.g., WACline offers three different format options to save an annotation, three navigation options that specify when the annotation is to be read, or two options that the user can choose from when storing the annotation in remote servers. Briefly, WACline accounts for 85 features, of which 25 are mandatory and 60 are optional. The feature diagram of WACline is presented in Fig. \ref{fig:featurediagram}. From this feature diagram we can calculate that WACline can yield a set of 5.77{*}10\textasciicircum 13 different products by combining all the configuration options, and obeying the 21 restrictions defined for the SPL. Despite its small size for an SPL, mastering WACline involves understanding over a hundred different notions and their inter-dependencies.

\section{SPL Cmaps: Concept Maps for SPLs} \label{Sec_SPL_CMaps}

This section presents \textit{SPL Cmaps} (Concept Maps for SPLs), our sensemaking scaffold proposal for domain understanding in SPLs. Traditional concept maps are self-contained, and hence, they can be assimilated to static documentation. By contrast, SPL Cmaps fulfil the third sensemaking guideline (see Section \ref{sec:SensemakingConceptMaps}) by enhancing concept maps with links. This turns SPL Cmaps into hyper-documents where links relate to other artefacts of the SPL. Adding links to SPL Cmaps allows newcomers to explore different representations of the SPL at their own pace by navigating the concept map, the features and their related core assets, be them code, requirements, architecture and so on.  Thus, SPL Cmaps help the newcomer understand the SPL domain by (1) presenting the set of most relevant concepts, understood by practitioners and that illustrate the features implemented in the SPL,  (2) tracing these concepts to their related features and core assets, hence showing the newcomer how variability is realised in that particular SPL and equipping them with the knowledge to build products in the area. To this end, SPL Cmaps rely on two types of traces: (1) concept-feature traces, created and evolved inside the concept map, and (2) feature-asset traces, managed in terms of embedded annotations. For annotative SPLs, maintaining these annotations has no additional effort, whereas in other scenarios it has been proven that the cost of creating and maintaining the annotations is negligible \cite{JiBAC15}. In addition, SPL Cmaps become interactive, a key feature to make learning more enjoyable and effective \cite{Lee06}. 

\begin{figure*}[h!]
  \centering
      \includegraphics[width=0.85\textwidth]{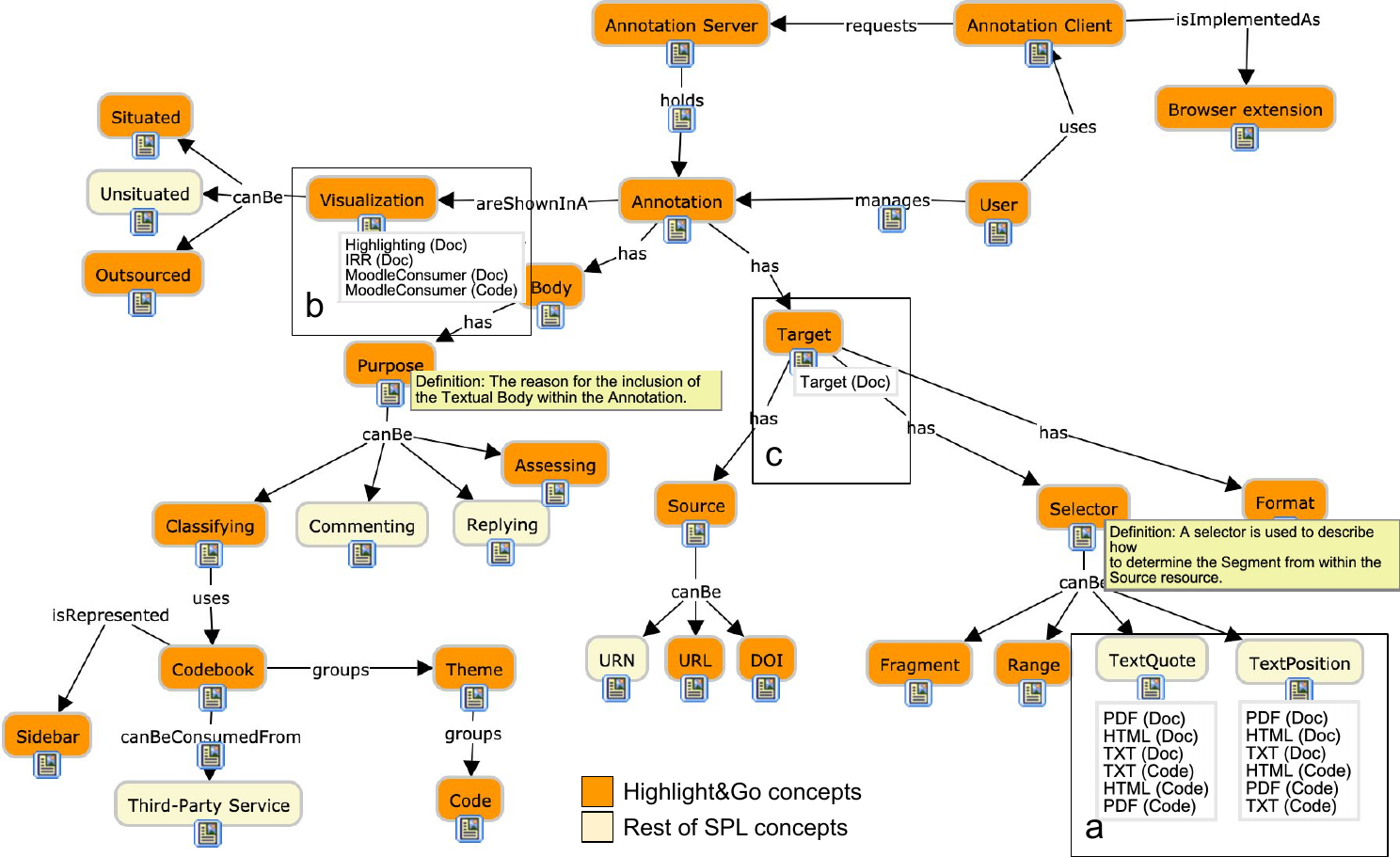}
  \caption{SPL Cmap of WACline  (partial view). }
  \label{fig:conceptmap}
\end{figure*}

\begin{figure*}[t!]
   \centering
      \includegraphics[width=\textwidth]{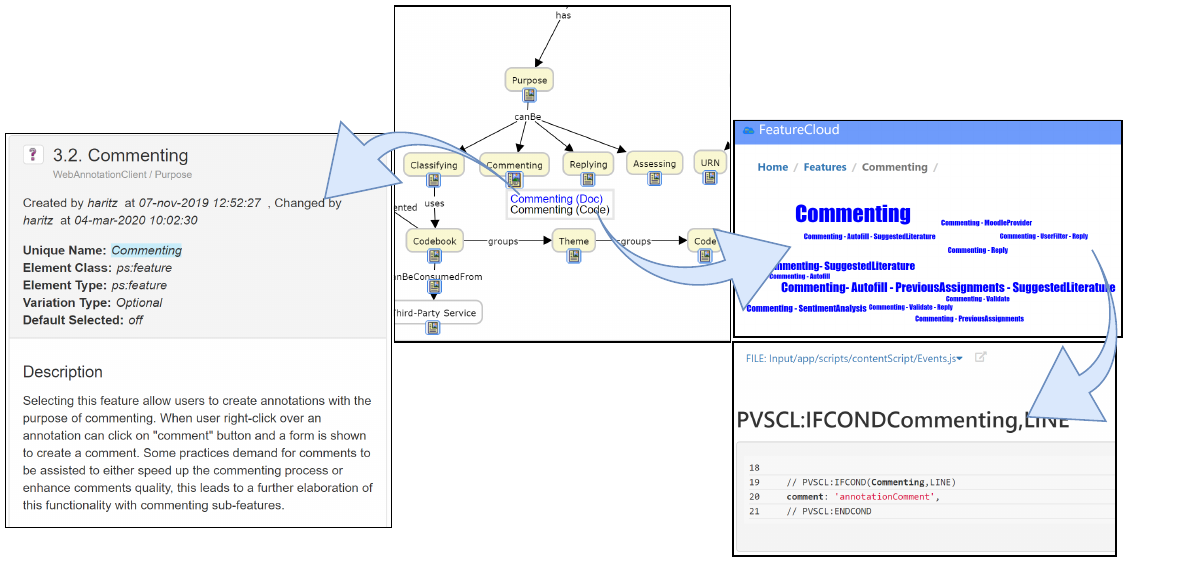}
  \caption{SPL Cmaps as hubs: understanding \textit{Commenting} concept}
  \label{fig:cmap-hub}
\end{figure*}

SPL Cmaps play two major roles: domain learning support and exploration hub. They are described with the help of WACline SPL (see Section \ref{Sec_wacline}).

\textit{\textbf{Domain learning support}} (see Fig. \ref{fig:conceptmap}). SPL Cmaps present the main concepts of the domain. However, these concepts need to be connected to the knowledge the newcomer already has. As an example, consider Jane, a developer that has previously worked in the application engineering team of \textit{Highlight\&Go}  \cite{Diaz2019HighlightAndGo}, one of the products derived from WACline. The main concepts from this application, which are already known to Jane, are highlighted in the SPL Cmap of Fig. \ref{fig:conceptmap} with an orange background. During Jane's onboarding journey, the rest of the SPL CMap concepts are presented (shown in Fig. \ref{fig:conceptmap} with a yellow background), thus connecting Jane's knowledge with the rest of the domain. As another aid in this endeavour, the SPL Cmap also exhibits tooltips with definitions used as glossary elements (see Fig. \ref{fig:conceptmap}).

\textit{\textbf{Exploration hub}} (see Fig. \ref{fig:cmap-hub}). Apart from concepts and relationships, SPL Cmaps present links between concepts and features. Links can be of different sorts:  many-to-many (e.g. in concepts \textit{TextQuote} and \textit{TextPosition}, with features \textit{PDF, TXT} and \textit{HTML} (see Fig. \ref{fig:conceptmap} (a)); one-to-many (e.g., in concept \textit{Visualization}  with \textit{Highlighting}, \textit{IRR} and \textit{MoodleConsumer} features (Fig. \ref{fig:conceptmap} (b)); or one-to-one (e.g. between \textit{Target} feature and concept (Fig. \ref{fig:conceptmap} (c)). Back to Jane, once she has set her focus on the \textit{Commenting} concept, she may right click for the related links to show up (see Fig. \ref{fig:cmap-hub}). Jane can now move to the documentation  (left-side window) or rather explore the tangling and scattering relationships that this feature holds with other features (right-side window). Specifically, clicking on the \textit{Commenting(Doc)} link, it transfers Jane to a web page that describes the \textit{Commenting} feature. Clicking on \textit{Commenting(Code)}, it moves Jane to WACline's \textit{FeatureCloud} instance \cite{DiazMM19}, where she can explore the implementation of \textit{Commenting} by analysing in which variation points this feature appears. 

Support for domain learning and the exploration hub can make the SPL CMaps a useful mechanism during the onboarding process. However, this vision clashes with the complexity of SPLs. Creating and maintaining such a hyper-document up to date is hardly feasible without assistance. This moves us to the next section.

\section{A Process for the Creation of SPL Cmaps}\label{Sec_construction}

\begin{figure*}[t!]
  \centering
      \includegraphics[width=0.9\textwidth]{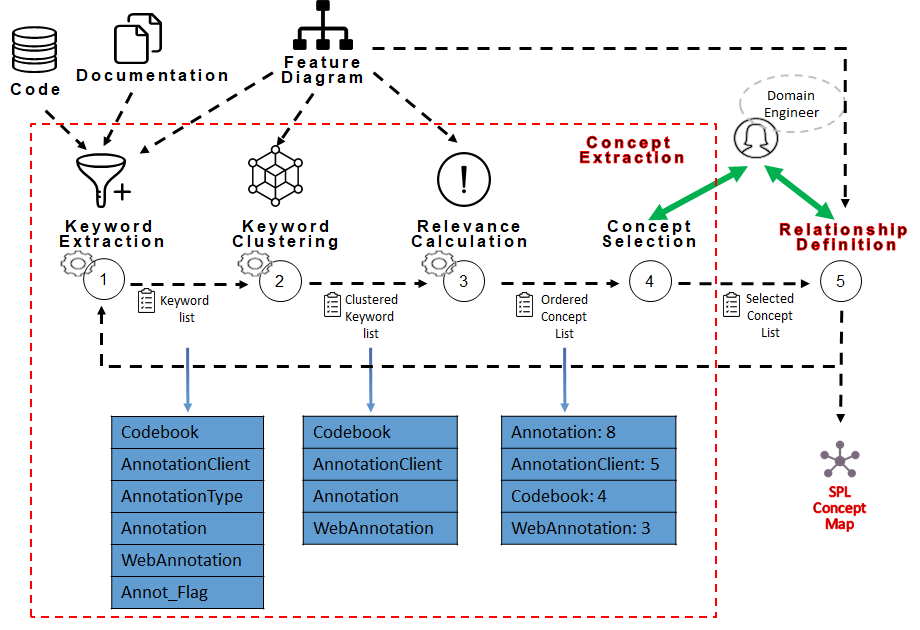}
  \caption{SPL Cmap creation process}
  \label{fig:creationprocess}
\end{figure*}

SPL analysis is normally conducted around the notion of \textit{feature}. Scope, product configuration, trace and other SPL concerns pivot around the notion of feature. It then makes sense to consider features also central throughout the building of SPL Cmaps. 

Fig. \ref{fig:creationprocess} summarises the SPL Cmap creation process. Here, domain engineers depart from the core assets of the SPL, including its feature diagram, requirements, documentation, source codebase and so on, to obtain the corresponding SPL Cmap. Specifically, our prototype has been developed using the feature diagram, the source code, and textual requirement documentation. This process comprises two consecutive subprocesses: \textit{Concept Extraction} and \textit{Relationship Definition}. 

\subsection{Concept Extraction}

The first step is to decide which concepts will conform the SPL Cmap. Following good practices \cite{Novak08}, concept maps should represent concepts in a hierarchical fashion, with the most inclusive (i.e. most general) concepts at the top of the map, and the most specific (i.e. less general) ones arranged hierarchically below. Using features as the guiding principle, we propose to start by extracting the concepts necessary to understand the features at the top level of the feature diagram, and then, to move down along its hierarchical levels, on the understanding that the parent feature is required to understand its children. 

The analysis then starts from the first-level features. Within each level, mandatory features are considered first\footnote{For a F feature at level L, its mandatory features at next level (L+1) are those present in all product variants where F feature is present.}. Mandatory features at first level are present in all product variants of the product line, hence, they constitute its core. This starting point concurs with the works of both Rajlich and Wilde \cite{Rajlich02} and Kr{\"u}ger et al. \cite{Krueger17} who start analysing common code and then, variable code in an incremental way. It is worth mentioning that, while desirable \cite{Krueger17}, not all SPL development paradigms maintain the trace between mandatory features and their realising core assets. In such cases, our approach would begin by processing all mandatory core assets together. 

Next, if needed, or-exclusive, or-inclusive and optional features are considered\footnote{These types of features define the variability on the SPL feature diagram, i.e. not all products of the SPL will have all those features; during the product configuration phase the application engineer will decide whether to select them or not.}. As shown in Fig. \ref{fig:creationprocess}, \textit{Concept Extraction} encompasses four main steps: \textit{Keyword Extraction}, \textit{Keyword Clustering}, \textit{Relevance Calculation} and \textit{Concept Selection}. 

\textit{\textbf{Keyword Extraction}}. This phase takes the SPL's core assets related to the analysed features as input, and yields a list of keywords. These keywords are selected by their use in the SPL documentation or its use through the implementation. This list is then enriched with information of the trace between the keywords and their related core assets, where they appear. If conducted manually, this process might be time-intensive for domain engineers. Assistance is needed. To this end, we resort to Natural Language Processing and Information Retrieval techniques \cite{Valerio06}. These techniques are successfully used to automatically extract information from text documents. In particular keyword extraction techniques, that is, the task of finding the words that best describe the subject of a text \cite{AbilhoaC14}, are useful in this scenario. Specifically, we use the \textit{linguakit} service\footnote{Available at: \url{https://linguakit.com/en/keyword-extractor}}.

\textit{\textbf{Keyword Clustering}}. This phase receives the previous keyword list as input and clusters related keywords into concepts \cite{Wartena08}. It can be carried out using available tools for machine learning, such as Weka\footnote{Available at: \url{https://www.cs.waikato.ac.nz/ml/weka/index.html}} or scikit\footnote{Available at: \url{https://scikit-learn.org/stable/}}.

\textit{\textbf{Relevance Calculation}}. This phase receives the clustered concept list as input and sorts it using the \textit{relevance of a concept} in the SPL core assets metric. Based on the TF-IDF metric \cite{Manning08}, the relevance value (\textit{rv}) metric considers not only the global frequency of a concept along all the SPL core assets, but also other criteria: how many different features the concept appears in, how many different core assets it appears in, the diversity of core asset types, i.e. documentation, implementation, and so on. Hence, for two concepts with the same global frequency, the one which appears in more features will have a bigger \textit{rv} value, and it will be higher in the concept list, which is the outcome of this process. More precisely, let C, F and A be the number of concepts, features, and core assets, respectively, let AT be the number of different core asset types, and let f(c), a(c) and at(c) be, respectively, the number of features, core assets and core asset types where the c concept appears. We represent the relevance of a \textit{c} concept with the \textit{rv(c)} value, calculated by multiplying four values (\(rv(c) = CF(c)*FF(c)*AF(c)*AD(c)\)), where:
\begin{itemize}
\item CF(c), the \textit{Concept Frequency}, is the number of occurrences of the c concept relative to the number of concepts: \(CF(c) = occurrences(c) / C\).
\item FF(c), the \textit{Feature Frequency}, is the number of features where the c concept appears relative to the number of features: \(FF(c) = f(c) / F\).
\item AF(c), the \textit{Core Asset Frequency}, is the number of core assets where the c concept appears relative to the number of core assets: \(AF(c) = a(c) / A\).
\item AD(c), the \textit{Core Asset Diversity}, is the number of different core asset types where the c concept appears relative to the number of different core asset types: \(AD(c) = at(c) / AT\).
\end{itemize}

The outcome is a concept list, where concepts are ordered by their relevance value (\textit{rv}). Moreover, this list is enriched with trace information, as each concept is accompanied by the core assets to which it is related. 

\textit{\textbf{Concept Selection}}. Throughout this manual process, the domain engineer will have to apply her knowledge to establish which concepts from the previous list should be prioritised to be shown in the SPL Cmap. 

\subsection{Relationship Definition}

Once the concepts have been extracted, the relationships are set by the domain engineer. As an aid in this process, we propose three guidelines. Again features will drive the process. The relationship between concepts and features has been defined in the concept extraction process. Set intersections will then guide the relationships between concepts. Specifically, let \(S_a\) and \(S_b\) be the sets of features related to concepts \(C_a\) and \(C_b\), respectively. The following can be applied next: 
\begin{itemize} 
\item If \(S_b\) is subset of \(S_a\), then \(C_b\) can be subsumed by \(C_a\) concept, and thus \(C_b\) is not added to the Cmap.
\item If there are elements in the intersection between \(S_b\) and \(S_a\), then both concepts will appear in the Cmap with a connection between them, the decision about its direction is left to the engineer.
\item If the intersection among \(S_b\) and \(S_a\) is empty, then both concepts will be shown in the Cmap, and another analysis step can be performed using the feature diagram:
\begin{itemize} 
\item If features in both sets are disconnected in the feature diagram, then no relationship will be drawn.
\item If features have a dependency relationship in the feature diagram, then a relationship will be set in the Cmap.
\end{itemize}
\end{itemize}

These two main processes yield a first draft of the concept map. They will be repeated for every level of the feature diagram until the domain engineer considers that the SPL Cmap content gathers sufficient concepts to show the fundamental aspects or concerns of the SPL for newcomers to be able to understand them.

\subsection{An SPL Cmap for \textit{WACline}: step-by-step construction\label{Wacline-step-by-step}}

%Revise the figure placement at the end
\begin{figure}
  \centering
      \includegraphics[width=0.5\textwidth]{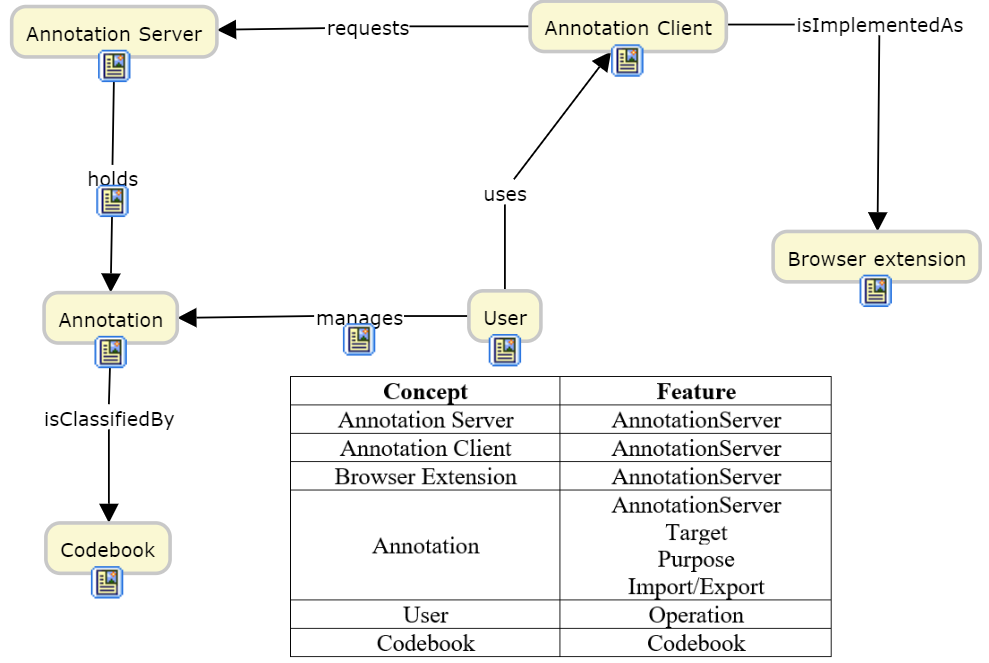}
  \caption{SPL Cmap (first-level features) and the tracing table counterpart.}
  \label{fig:conceptmap1level}
\end{figure}

Next paragraphs illustrate how WACline's Cmap can be gradually obtained, following the aforementioned processes. Tables in blue of Fig. \ref{fig:creationprocess} show simplified excerpts of the lists that conform the outcome of the following steps.

\textit{Keyword Extraction}. WACline's feature diagram (see Fig. \ref{fig:featurediagram}) provides a first functional characterisation in terms of first-level features: \textit{AnnotationServer}, \textit{Target}, \textit{Codebook}, \textit{Purpose}, \textit{Operation}, \textit{ImportExport}. The latter is optional  while the rest  are mandatory. In the first step, throughout the core assets of those features, keywords such as \textit{AnnotationType}, \textit{Annot\_Flag}, \textit{Annotation}, \textit{AnnotationList}, \textit{AnnotationClient}, are extracted. 

\textit{Keyword Clustering}. Using the clustering algorithm, related keywords are merged together creating a first list of concepts. In the example, some of these concepts are \textit{Annotation}, \textit{Codebook}, \textit{AnnotationClient}, \textit{AnnotationServer}, \textit{Web Annotation} and \textit{Web Browser}. For example, the \textit{Annotation} concept brings together the \textit{AnnotationType}, \textit{Annot\_Flag} and \textit{Annotation} keywords, among others. 

\textit{Relevance Calculation}. This step will reorder those concepts depending on their relative importance within the core assets. Thus, in the example, \textit{Annotation} is the most relevant concept.

\textit{Concept Selection}. Next, the domain engineer might decide that \textit{Annotation} and \textit{Web Annotation} stand for the same idea, or that the \textit{User} concept is required, even though it obtains a low relevance value (not shown in the excerpt), because this concept accounts for the stakeholder who manages annotations (i.e. who does operations with annotations, and \textit{CRUD operation} appears in the list). Thus, those selected concepts are shown in Fig. \ref{fig:conceptmap1level}.

\textit{Relationship Definition}. Next, the domain engineer has to state the relationships between concepts using the aforementioned guidelines as an aid. It is up to them to decide the direction and name of each relationship. The resulting SPL Cmap is displayed in Fig. \ref{fig:conceptmap1level}. This figure also shows a table with the trace between each concept and the features it is related to. The trace information is carried throughout the steps from the keyword extraction, and it is enacted in the SPL Cmap as links, as explained in Section \ref{Sec_SPL_CMaps}.

\textit{Repeat if required}. Once the processes are finished for first-level features, the domain engineer can apply them to second-level features, such as, \textit{Format}, \textit{Source}, \textit{Create}, \textit{Read}, \textit{Classifying}, \textit{Import}, and so on. The engineer might decide to halt the process at any time, when they consider that the current SPL Cmap is accurate enough. Sometimes the domain engineer will have to overturn a previous decision. For instance, in the case of WACline it is not until the second level, i.e. when the \textit{Classifying} feature is processed, that the domain engineer decides that the \textit{Classifying} concept is needed and that the \textit{Codebook} concept is related to it, instead of to the \textit{Annotation} concept (i.e. the former decision). Fig. \ref{fig:conceptmap} shows a partial view of the final SPL Cmap, that contains the key concepts and relationships, and the links to core assets which are necessary to understand WACline, according to the domain engineer's knowledge.

\section{Evaluation}\label{Sect_Evaluation}
\begin{table*}[]
\centering
\resizebox{0.95\textwidth}{!}{
\small
\begin{tabular}{|l|l|ccccc|c|c|}
\hline
    & {\color[HTML]{000000} \textbf{}}                                                                                                 & {\color[HTML]{000000} \textbf{P1}} & {\color[HTML]{000000} \textbf{P2}} & {\color[HTML]{000000} \textbf{P3}} & {\color[HTML]{000000} \textbf{P4}} & \textbf{P5} & {\color[HTML]{000000} \textbf{Avg.}} & {\color[HTML]{000000} \textbf{ST. Dev}} \\ \hline
Q1  & {\color[HTML]{000000} In general, being able to do your assigned work with WACline was easy/straightforward}                     & {\color[HTML]{000000} 2}           & {\color[HTML]{000000} 4}           & {\color[HTML]{000000} 3}           & {\color[HTML]{000000} 4}           & 2           & {\color[HTML]{000000} 3}                & {\color[HTML]{000000} 1}                \\ \hline
Q2  & {\color[HTML]{000000} In general, being able to understand the feature model of WACline was easy/straightforward}                & {\color[HTML]{000000} 3}           & {\color[HTML]{000000} 4}           & {\color[HTML]{000000} 4}           & {\color[HTML]{000000} 5}           & 3           & {\color[HTML]{000000} 3.8}              & {\color[HTML]{000000} 0.84}             \\ \hline
Q3  & {\color[HTML]{000000} In general, being able to understand the code of WACline was easy/straightforward}                         & {\color[HTML]{000000} 3}           & {\color[HTML]{000000} 4}           & {\color[HTML]{000000} 2}           & {\color[HTML]{000000} 2}           & 2           & {\color[HTML]{000000} 2.6}              & {\color[HTML]{000000} 0.89}             \\ \hline
Q4  & {\color[HTML]{000000} I found  features themselves easy/straightforward to understand.}                                          & {\color[HTML]{000000} 4}           & {\color[HTML]{000000} 3}           & {\color[HTML]{000000} 5}           & {\color[HTML]{000000} 4}           & 3           & {\color[HTML]{000000} 3.8}              & {\color[HTML]{000000} 0.82}             \\ \hline
Q5  & {\color[HTML]{000000} I found interrelationships among features and their impact in the SPL easy/straightforward to understand.} & {\color[HTML]{000000} 2}           & {\color[HTML]{000000} 5}           & {\color[HTML]{000000} 4}           & {\color[HTML]{000000} 4}           & 2           & {\color[HTML]{000000} 3.4}              & {\color[HTML]{000000} 1.26}             \\ \hline
Q6  & {\color[HTML]{000000} In my opinion, understading the domain first is important to understand the WACline code.}                 & {\color[HTML]{000000} 5}           & {\color[HTML]{000000} 2}           & {\color[HTML]{000000} 5}           & {\color[HTML]{000000} 5}           & 5           & {\color[HTML]{000000} 4.4}              & {\color[HTML]{000000} 1.50}             \\ \hline
Q7  & {\color[HTML]{000000} In my opinion, understanding feature code blocks is important to understand the WACline code.}             & {\color[HTML]{000000} 5}           & {\color[HTML]{000000} 3}           & {\color[HTML]{000000} 4}           & {\color[HTML]{000000} 4}           & 4           & {\color[HTML]{000000} 4}                & {\color[HTML]{000000} 0.82}             \\ \hline
Q8  & {\color[HTML]{000000} In my opinion,  understanding features is important to understand feature code blocks.}                    & {\color[HTML]{000000} 5}           & {\color[HTML]{000000} 2}           & {\color[HTML]{000000} 5}           & {\color[HTML]{000000} 2}           & 4           & {\color[HTML]{000000} 3.6}              & {\color[HTML]{000000} 1.73}             \\ \hline
Q10 & {\color[HTML]{000000} I found locating the implementation of each feature easy/straightforward.}                                 & {\color[HTML]{000000} 1}           & {\color[HTML]{000000} 2}           & {\color[HTML]{000000} 2}           & {\color[HTML]{000000} 2}           & 3           & {\color[HTML]{000000} 2}                & {\color[HTML]{000000} 0.71}             \\ \hline
\end{tabular}
}
\normalsize
\caption{Participants' perception after working with WACline}
\label{tab:onboarding-factors}
\end{table*}

This work presents SPL Cmaps as an interactive way for newcomers to explore the SPL. The aim is to ease onboarding by broadening their domain knowledge and reducing information overload. WACline was used as a running example. Hence, two questions remain unanswered:  (1) are SPL Cmaps useful to ease onboarding when application engineers are incorporated into the domain engineering team? and, with the aim to minimise frustration, (2) are SPL Cmaps easy to use? This section tackles these questions with a formative evaluation.   

\textit{\textbf{Participants}}. We recruited five participants that were originally involved in the customisation of \textit{Hightlight\&Go}, a product derived from WACline and then proceeded to work with the whole SPL. They all worked approximately 300 hours in WACline. Participants were recent graduates who hold a Computer Science Bachelor degree at the University of the Basque Country (UPV/EHU). They all had comparable previous experience in SPLs, about four weeks in an advanced software engineering undergraduate course and no previous experience in WACline's domain before they started their projects. Our proposal is directed to SPL practitioners and using newly graduates as participants can be seen as at least controversial \cite{Feldt18}. In this particular case, given the participants' backgrounds, we believe that they can be assimilated to application engineers being incorporated into an SPL domain engineering team. 

\textit{\textbf{Method}}. We used a survey methodology. Two questionnaires were delivered. One gathers participants' onboarding experience with WACline using LIKERT-scale rated questions with ranges from 1 ("Strongly disagree") to 5 ("Strongly agree"). The questions were divided into three groups: (1) Q1-Q3 gather general insights about their experience working with WACline, (2) Q4 and Q5 refer to complexity of understanding the feature diagram, and (3) Q6-Q9 try to summarise their opinion about WACline's code (see Table  \ref{tab:onboarding-factors}). The second questionnaire resorts to the \textit{Technology Acceptance Model (TAM)} to collect  the perceived \textit{usefulness} and \textit{ease of use} of SPL Cmaps as an instrument for understanding an SPL domain and its feature diagram (see Table \ref{tab:TAM}). We also added some open questions to both questionnaires in case participants wanted to add comments.

\textit{\textbf{Procedure}}. Participants were contacted by email explaining the evaluation procedure and that the results would be completely anonymous. The email message asked them to perform the evaluation in one session for the sake of avoiding maturation. They first answered the questionnaire of their onboarding experience. Next, we asked them to explore the SPL Cmap of WACline and to navigate the links to the documentation and code. They then answered the TAM questionnaire.

\begin{table*}[]
\centering
\resizebox{0.95\textwidth}{!}{%
\small
\begin{tabular}{lccccccc}
\hline
\multicolumn{1}{|l|}{{\color[HTML]{000000} \textbf{Usefulness}}}                                                                  & \multicolumn{1}{l}{{\color[HTML]{000000} \textbf{P1}}} & \multicolumn{1}{l}{{\color[HTML]{000000} \textbf{P2}}} & \multicolumn{1}{l}{{\color[HTML]{000000} \textbf{P3}}} & \multicolumn{1}{l}{{\color[HTML]{000000} \textbf{P4}}} & \multicolumn{1}{l|}{\textbf{P5}} & \multicolumn{1}{l|}{{\color[HTML]{000000} \textbf{Avg.}}} & \multicolumn{1}{l|}{{\color[HTML]{000000} \textbf{ST. Dev}}} \\ \hline
\multicolumn{1}{|l|}{{\color[HTML]{000000} Using Concept Maps in my job would have enabled me to accomplish tasks more quickly.}} & {\color[HTML]{000000} 5}                               & {\color[HTML]{000000} 4}                               & {\color[HTML]{000000} 5}                               & {\color[HTML]{000000} 7}                               & \multicolumn{1}{c|}{5}           & \multicolumn{1}{c|}{{\color[HTML]{000000} 5.2}}              & \multicolumn{1}{c|}{{\color[HTML]{000000} 1.09}}             \\ \hline
\multicolumn{1}{|l|}{{\color[HTML]{000000} Using Concept Maps in my job would increase my productivity.}}                         & {\color[HTML]{000000} 6}                               & {\color[HTML]{000000} 5}                               & {\color[HTML]{000000} 5}                               & {\color[HTML]{000000} 7}                               & \multicolumn{1}{c|}{6}           & \multicolumn{1}{c|}{{\color[HTML]{000000} 5.8}}              & \multicolumn{1}{c|}{{\color[HTML]{000000} 0.83}}             \\ \hline
\multicolumn{1}{|l|}{{\color[HTML]{000000} Using Concept Maps would have improved my job performance.}}                           & {\color[HTML]{000000} 5}                               & {\color[HTML]{000000} 4}                               & {\color[HTML]{000000} 4}                               & {\color[HTML]{000000} 6}                               & \multicolumn{1}{c|}{5}           & \multicolumn{1}{c|}{{\color[HTML]{000000} 4.8}}              & \multicolumn{1}{c|}{{\color[HTML]{000000} 0.83}}             \\ \hline
\multicolumn{1}{|l|}{{\color[HTML]{000000} Using Concept Maps would have enhanced my effectiveness on the job.}}                  & {\color[HTML]{000000} 5}                               & {\color[HTML]{000000} 3}                               & {\color[HTML]{000000} 4}                               & {\color[HTML]{000000} 6}                               & \multicolumn{1}{c|}{6}           & \multicolumn{1}{c|}{{\color[HTML]{000000} 4.8}}              & \multicolumn{1}{c|}{{\color[HTML]{000000} 1.3}}              \\ \hline
\multicolumn{1}{|l|}{{\color[HTML]{000000} Using Concept Maps would have made it easier to do my job.}}                           & {\color[HTML]{000000} 5}                               & {\color[HTML]{000000} 5}                               & {\color[HTML]{000000} 5}                               & {\color[HTML]{000000} 6}                               & \multicolumn{1}{c|}{5}           & \multicolumn{1}{c|}{{\color[HTML]{000000} 5.2}}              & \multicolumn{1}{c|}{{\color[HTML]{000000} 0.44}}             \\ \hline
\multicolumn{1}{|l|}{{\color[HTML]{000000} I would have found Concept Maps useful in my job.}}                                    & {\color[HTML]{000000} 5}                               & {\color[HTML]{000000} 5}                               & {\color[HTML]{000000} 6}                               & {\color[HTML]{000000} 7}                               & \multicolumn{1}{c|}{5}           & \multicolumn{1}{c|}{{\color[HTML]{000000} 5.6}}              & \multicolumn{1}{c|}{{\color[HTML]{000000} 0.89}}            \\ \hline
\multicolumn{1}{|l|}{{\color[HTML]{000000} \textbf{Avg.}}}                                                                     & {\color[HTML]{000000} 5.17}                            & {\color[HTML]{000000} 4.33}                            & {\color[HTML]{000000} 4.83}                            & {\color[HTML]{000000} 6.50}                            & \multicolumn{1}{c|}{5.33}        & \multicolumn{1}{c|}{{\color[HTML]{000000} 5.23}}             & \multicolumn{1}{c|}{{\color[HTML]{000000} 0.8}}              \\ \hline
{\color[HTML]{000000} }                                                                                                           & {\color[HTML]{000000} }                                & {\color[HTML]{000000} }                                & {\color[HTML]{000000} }                                & {\color[HTML]{000000} }                                &                                  & {\color[HTML]{000000} }                                      & {\color[HTML]{000000} }                                      \\ \hline
\multicolumn{1}{|l|}{{\color[HTML]{000000} \textbf{Ease of use}}}                                                                 & {\color[HTML]{000000} \textbf{P1}}                     & {\color[HTML]{000000} \textbf{P2}}                     & {\color[HTML]{000000} \textbf{P3}}                     & {\color[HTML]{000000} \textbf{P4}}                     & \multicolumn{1}{c|}{\textbf{P5}} & \multicolumn{1}{c|}{{\color[HTML]{000000} \textbf{Avg.}}} & \multicolumn{1}{c|}{{\color[HTML]{000000} \textbf{ST. Dev}}}          \\ \hline
\multicolumn{1}{|l|}{{\color[HTML]{000000} Learning to operate Concept Maps would have been easy for me.}}                        & {\color[HTML]{000000} 5}                               & {\color[HTML]{000000} 7}                               & {\color[HTML]{000000} 7}                               & {\color[HTML]{000000} 6}                               & \multicolumn{1}{c|}{4}           & \multicolumn{1}{c|}{{\color[HTML]{000000} 5.8}}              & \multicolumn{1}{c|}{{\color[HTML]{000000} 1.3}}              \\ \hline
\multicolumn{1}{|l|}{{\color[HTML]{000000} I would have found it easy to get Concept Maps to do what I want it to do.}}           & {\color[HTML]{000000} 5}                               & {\color[HTML]{000000} 5}                               & {\color[HTML]{000000} 6}                               & {\color[HTML]{000000} 6}                               & \multicolumn{1}{c|}{5}           & \multicolumn{1}{c|}{{\color[HTML]{000000} 5.4}}              & \multicolumn{1}{c|}{{\color[HTML]{000000} 0.54}}             \\ \hline
\multicolumn{1}{|l|}{{\color[HTML]{000000} My interaction with Concept Maps would have been clear and understandable.}}           & {\color[HTML]{000000} 5}                               & {\color[HTML]{000000} 6}                               & {\color[HTML]{000000} 7}                               & {\color[HTML]{000000} 6}                               & \multicolumn{1}{c|}{5}           & \multicolumn{1}{c|}{{\color[HTML]{000000} 5.8}}              & \multicolumn{1}{c|}{{\color[HTML]{000000} 0.83}}             \\ \hline
\multicolumn{1}{|l|}{{\color[HTML]{000000} I would have found Concept Maps to be flexible to interact with.}}                     & {\color[HTML]{000000} 5}                               & {\color[HTML]{000000} 4}                               & {\color[HTML]{000000} 6}                               & {\color[HTML]{000000} 6}                               & \multicolumn{1}{c|}{5}           & \multicolumn{1}{c|}{{\color[HTML]{000000} 5.2}}              & \multicolumn{1}{c|}{{\color[HTML]{000000} 0.83}}             \\ \hline
\multicolumn{1}{|l|}{{\color[HTML]{000000} It would have been easy for me to become skillful at using Concept Maps.}}             & {\color[HTML]{000000} 5}                               & {\color[HTML]{000000} 5}                               & {\color[HTML]{000000} 5}                               & {\color[HTML]{000000} 5}                               & \multicolumn{1}{c|}{6}           & \multicolumn{1}{c|}{{\color[HTML]{000000} 5.2}}              & \multicolumn{1}{c|}{{\color[HTML]{000000} 0.44}}             \\ \hline
\multicolumn{1}{|l|}{{\color[HTML]{000000} I would have found Concept Maps easy to use.}}                                         & {\color[HTML]{000000} 5}                               & {\color[HTML]{000000} 7}                               & {\color[HTML]{000000} 7}                               & {\color[HTML]{000000} 6}                               & \multicolumn{1}{c|}{6}           & \multicolumn{1}{c|}{{\color[HTML]{000000} 6.2}}              & \multicolumn{1}{c|}{{\color[HTML]{000000} 0.83}}             \\ \hline
\multicolumn{1}{|l|}{{\color[HTML]{000000} \textbf{Avg.}}}                                                                     & {\color[HTML]{000000} 5}                               & {\color[HTML]{000000} 5.67}                            & {\color[HTML]{000000} 6.33}                            & {\color[HTML]{000000} 5.83}                            & \multicolumn{1}{c|}{5.16}        & \multicolumn{1}{c|}{{\color[HTML]{000000} 5.6}}              & \multicolumn{1}{c|}{{\color[HTML]{000000} 0.53}}             \\ \hline
\end{tabular}
}

\normalsize

\caption{Perceived \textit{usefulness} and \textit{ease of use} of SPL Cmaps using Davis' scales \cite{Davis89}}
\label{tab:TAM}
\end{table*}
\normalsize

\textit{\textbf{Results}} (see Table \ref{tab:onboarding-factors}).  If we dive into the first group of questions (Q1-Q3), we can see that participants did not perceive any task as particularly difficult, but neither they found them too simple, results range from 2.6 to 3.8 on average. Understanding code is discerned as the most complex task, since three newcomers rated it with 2 and has a mean of 2.6. Moreover, the second group (i.e. Q4 and Q5) and Q2 are consistent: participants found the feature diagram and its dependencies quite understandable, although P1 and P5 had problems understanding the feature diagram and its interrelations. Finally, on the issue of code understanding several aspects can be discerned. Participants found feature code locating to be an arduous task. As one commented, \textit{"As some features are scattered over a group of files, following the course of some features along the code was a sort of a maze"}. When it comes to what is key in order to understand WACline's code, understanding the domain was perceived as the most crucial facet and it was rated with an average of 4.4. Regarding \textit{ease of use} and \textit{usefulness}, participants rank them both over 5 on average. 

Obtained results indicate that SPL Cmaps would have good acceptability for the task, but even if they are promising, this was a formative evaluation and therefore, a first test of our proposal. Thus, these results need to be interpreted with caution.  

\textbf{Threats to Validity}. \emph{Construct validity} refers to the degree of accuracy to which the variables defined in a study measure the constructs of interest. In this evaluation, we used two questionnaires, one created by us and the TAM. The latter has been broadly used. While the one we created to analyse problems with first SPL use requires more validation, its results concur with problems reported in the literature \cite{Acher17, Capilla13}. \emph{Internal validity} is concerned with the conduct of the study. A first concern is the number of participants. While these results are promising, larger evaluations are needed to confirm them. The evaluation was online, and was thus impossible to control thoroughly. Nonetheless, in real industrial settings, newcomers use and explore the onboarding tools unsupervised, which makes our setting more realistic \cite{Viviani19}. As for \emph{External validity}, i.e. the ability to generalise the results of our experiment to other settings, this study's newcomers are recent graduates with a shallow understanding of SPLs and pure::variants (i.e. the variability management tool). Results might vary for more mature practitioners. Besides practitioners, the SPL itself might have an impact on the results. WACline is a relatively small SPL. Yet, the need for some SPL-introduction road-map becomes even more evident the larger the SPL is. A concern is that of the scalability of the SPL Cmap. Large SPLs will likely result in huge maps. That said, "perspectives" can be defined whereby the SPL Cmap focuses on an area of the SPL where it is more likely newcomers will be incorporated.
%At this point, we believe that our results can be generalise to middle size SPL companies suffering the onboarding problem. 

\section{Conclusions and Future Work}
This work presents SPL Cmaps as sensemaking scaffolds for newcomers into SPL domain engineering. SPL Cmaps are built around the semantics of the discipline, depart from knowledge newcomers already have while permitting newcomers explore different representations of the SPL by the means of traces (links). In this way, newcomers are introduced into the SPL domain by exploring both the concepts practitioners use and the variability that yields the different products of the SPL.  

This work elaborates on the construction of SPL Cmaps. Features are used as the guiding principle while the rest of the documentation is used as a  complement. On top, SPL Cmaps are enriched with links to external resources. In this way, SPL Cmaps become hubs to the rest of the SPL core assets, permitting newcomers to explore the different abstraction levels.  First insights are provided for WACline, an academic SPL. A first formative evaluation has been conducted. Results are encouraging yet threats to validity remain.

At this point, this proposal aims at keeping traces between concepts, features, their documentation and codebase. Nevertheless, as a future work, we plan to delve into the mapping between concepts and the rest of core assets of the SPL such as architecture, requirements or design. Our goal is to provide a complete approach that, starting from the most abstract concepts, aids the newcomer in exploring the ins and outs of the SPL. Regarding the feature-driven construction algorithm, it can be used by SPL domain engineers to create SPL Cmaps. It is a preliminary approach that applies a breadth-first search of the feature diagram, starting from mandatory features and then optional ones. More analysis is needed in order to evaluate whether other traversal orders would be more appropriate, or even whether the relevance calculation formula should be adjusted. An evaluation of the experience and the opinion of expert SPL engineers would be necessary. 

\textbf{\textit{Acknowledgements.}} This work is supported by Spanish Ministry of Science, Innovation and Universities grant number RTI2018-099818-B-I00 and the Ministry of Education with grant number MCIU-AEI TIN2017-90644-REDT (TASOVA). Our gratitude to Haritz Medina for his help with WACline and to the five participants in the evaluation for their time and comments.
%\section{Data Availability}
%CMap of the WACLine SPL: \url{https://rebrand.ly/wacline-cmap}

%% Depending on the space we have left, appendixes with the questionnaires used in the evaluation
\bibliographystyle{IEEEtran}
\bibliography{20CMap4SPL}

%\appendix

\end{document}